\documentclass[conference]{IEEEtran}
\usepackage{amsmath,amsfonts}
\usepackage{algorithmic}
\usepackage{algorithm}
\usepackage{array}
\usepackage[caption=false,font=normalsize,labelfont=sf,textfont=sf]{subfig}
\usepackage{textcomp}
\usepackage{stfloats}
\usepackage{url}
\usepackage{graphicx}
\usepackage{cite}
\usepackage{graphicx}
\usepackage{array}
\usepackage{multirow}
\usepackage{tabularx}

\usepackage{tikz}
\def\BibTeX{{\rm B\kern-.05em{\sc i\kern-.025em b}\kern-.08em
    T\kern-.1667em\lower.7ex\hbox{E}\kern-.125emX}}
\newcommand{\Cross}{\mathbin{\tikz [x=1.4ex,y=1.4ex,line width=.2ex] \draw (0,0) -- (1,1) (0,1) -- (1,0);}}
\begin{document}
\title{Impact of Weather on Satellite Communication: Evaluating Starlink  Resilience\\}
\author{\IEEEauthorblockN{
Muhammad Asad Ullah\IEEEauthorrefmark{1},
Antti Heikkinen\IEEEauthorrefmark{1},
Mikko Uitto\IEEEauthorrefmark{1},
Antti Anttonen\IEEEauthorrefmark{1},
Konstantin Mikhaylov\IEEEauthorrefmark{2}
}
\IEEEauthorblockA{\IEEEauthorrefmark{1}VTT Technical Research Centre of Finland Ltd., Espoo, Finland}
\IEEEauthorblockA{\IEEEauthorrefmark{2}Centre for Wireless Communications, University of Oulu, Oulu, Finland}}

\maketitle

\begin{abstract}
Satellite communications have emerged as one of the most feasible solutions to provide global wireless coverage and connect the unconnected. Starlink dominates the market with over 7,000 operational satellites in low Earth orbit (LEO) and offers global high-speed and low-latency Internet service for stationary and mobile use cases, including in-motion connectivity for vehicles, vessels, and aircraft. Starlink terminals are designed to handle extreme weather conditions. Starlink recommends a flat high performance~(FHP) terminal for users living in areas with extreme weather conditions. The earlier studies evaluated Starlink's FHP throughput for stationary and in-motion users without providing a detailed analysis of how weather affects its performance. There remains a need to investigate the impact of weather on FHP's throughput. In this paper, we address this shortcoming by analyzing the impact of weather on Starlink's performance in Oulu, Finland, a city located in Northern Europe near the Arctic Circle. Our measurements reveal that rain degrades median uplink and downlink throughput by  52.27\% and 37.84\%, respectively. On the contrary, there was no noticeable impact on the round-trip time. Additionally, we also examine the impact of cloud cover on the Starlink throughput. The linear regression analysis reveals the negative relationship between throughput and cloud cover. The cloud cover of up to 12.5\%  has around 20\% greater throughput than the cloud cover of 87.5\%.
\end{abstract}

\begin{IEEEkeywords}
Clouds, ESIM, LEO, satellite, rain, throughput.
\end{IEEEkeywords}

\section{Introduction}
\let\thefootnote\relax\footnote{This work has been accepted in the 2025 IEEE 101st Vehicular Technology Conference: VTC2025-Spring. Copyright has been transferred to IEEE.}The terrestrial network (TN) does not provide global coverage. The remote areas are still not benefiting from digital technologies due to the lack of wireless coverage. The coverage gap is even larger in the oceanic areas, which make up 70\% of the Earth's surface. This highlights a major digital divide, which is primarily due to high network infrastructure deployment costs and low investments in remote areas.  A non-terrestrial network (NTN) is a native component of the 6G for enabling global connectivity. NTN has the potential to deliver ubiquitous wireless coverage that will connect the unconnected~\cite{Kodheli}.

In recent years, significant technological and regulatory advancements have reshaped NTN systems, particularly the satellite communications sector. Today, more than 12,000 satellites are operational, with over 84\% located in low Earth orbit (LEO).  SpaceX, OneWeb, Iridium, and GlobalStar are the key players in LEO~\cite{Sandrine,Heli,Anastasia}, where SpaceX's Starlink dominates the market with over 7,000 operational satellites in~LEO.

Recent years have witnessed several measurements that investigated Starlink's stationary and in-motion performance~\cite{Dominic_2024,Sami,Beckman,Melisa,Bin,Zhao,Michel,Kassem,Mohan}. These measurements show that Starlink provides high-speed and low-latency Internet connectivity for both stationary and in-motion use. However, its throughput experiences fluctuations due to the dynamic nature of channel conditions over time. The work in~\cite{Michel} reports Starlink's user-perceived performance in Western Europe, evaluating latency, packet loss rate, and throughput. In~\cite{Kassem}, authors examine website browser performance using data from eighteen Starlink user terminals deployed across the world.  In~\cite{LENS}, thirteen Starlink terminals each with different hardware revisions, service subscriptions and sky obstruction ratios deployed across three continents are used to study the round-trip time~(RTT). Collectively, these studies~\cite{Michel, Kassem, Zhao,Sami,LENS,Mohan} reveal that Starlink can support demanding applications, including live streaming, video conferencing, and gaming.


To understand in-motion performance, the works in  \cite{Beckman} and \cite{Bin} compare the Starlink and cellular network throughput measured in the Sweden and USA, respectively. The comparison indicates that, on average, Starlink outperforms cellular networks. However, Starlink experiences more frequent connection outage.  In~\cite{Asad_Starlink,Dominic_2024}, throughput results are reported for stationary and in-motion scenarios in Finland and Germany, respectively. It was observed that the Starlink in-motion throughput is lower and experiences frequent outages than stationary. This is mainly due to the roadside blockages obstructing the line-of-sight between the terminal and satellite~\cite{Sami}.

Starlink uses Ku-band, as one option, specifically 14.0-14.5~GHz for uplink and 10.7-12.7~GHz for downlink communications ~\cite{Humphreys,nhan}. Uplink has a total spectrum of 500~MHz, which is further divided into eight channels with a bandwidth equivalent to 62.5~MHz. Downlink deploys 2~GHz spectrum that is equally divided into eight channels, each with a bandwidth of 250~MHz. In each uplink channel, the Starlink terminal's emissions toward the horizon are restricted to a maximum of -72.76~dBW/Hz, equivalent to the maximum equivalent isotropically radiated power (EIRP) of around 3.2~W. In Ku-band downlink, the highest EIRP density is -51.6~dBW/Hz for a satellite with an orbital height of 540~km. This makes the downlink signals stronger than the uplink. It is worth mentioning that Ku-band is vulnerable to weather, e.g., rain and humidity~\cite{Sudarshana}.

As of today, only a few studies have examined the impact of weather on Starlink performance. In \cite{Zhao}, it was observed that thunderstorms affect real-time multimedia services, causing video pauses and audio cut-offs. However, there was no significant difference in real-time multimedia services during rainy weather conditions. According to the measurements in Canada \cite{Sami,Sami2}, it is evident that the precipitation reduces the throughput on average by 27\%. During 4.1 to 5.2 mm of precipitation, the User Datagram Protocol (UDP) downlink throughput was dropped by almost 50\%. The measurements conducted in Germany and the Netherlands report that the UDP throughput decreased by 30\% with 4-5 mm of rain \cite{WetLinks,Dominic_3}. Clouds lower the downlink throughput by 5\%. Similarly, in \cite{Careau}, results from Sweden confirm Starlink's throughput decreases with rain. In \cite{Dominic_2}, measurements conducted under cloudy weather conditions show that UDP download throughput was reduced by more than 6 Mbps. However, no significant impact was observed on upload throughput. Both clouds and rain have a noticeable effect on the page transit time (PTT) \cite{Kassem}. Particularly, moderate rain has a higher PTT compared to overcast clouds or light rain.

\begin{table}[t]
    \centering
    \caption{List of the relevant rain-focused reviewed papers.}
  \resizebox{\columnwidth}{!}{\begin{tabular}{|l|c|c|c|c|c|c|}
        \hline
        \textbf{Ref.} & \textbf{Year}& \textbf{FHP}  & \textbf{Link} & \textbf{Protocol}&  \textbf{Latency} & \textbf{Region} \\
        \textbf{} & \textbf{}&\textbf{}& \textbf{} & \textbf{}  &  \textbf{} & \textbf{country}\\
        \hline
        \textbf{\cite{Sami}} & 2022&  $\Cross$  & DL, UL &TCP, UDP&$\Cross$& Canada\\
        \textbf{\cite{WetLinks}} & 2024 & $\Cross$ &   DL &UDP&$\Cross$&  Germany\\
    \textbf{\cite{Careau}} &2024& $\Cross$  & DL, UL &TCP, UDP&$\Cross$& Sweden\\

        \textbf{Our paper} & 2025&  \checkmark  & DL, UL &TCP&\checkmark& Finland\\
        \hline 
   \end{tabular}}
    \label{tab:TabI}

    \vspace{-10pt}
\end{table}

\begin{figure}[t!]
\centerline{\includegraphics*[width=0.5\textwidth]{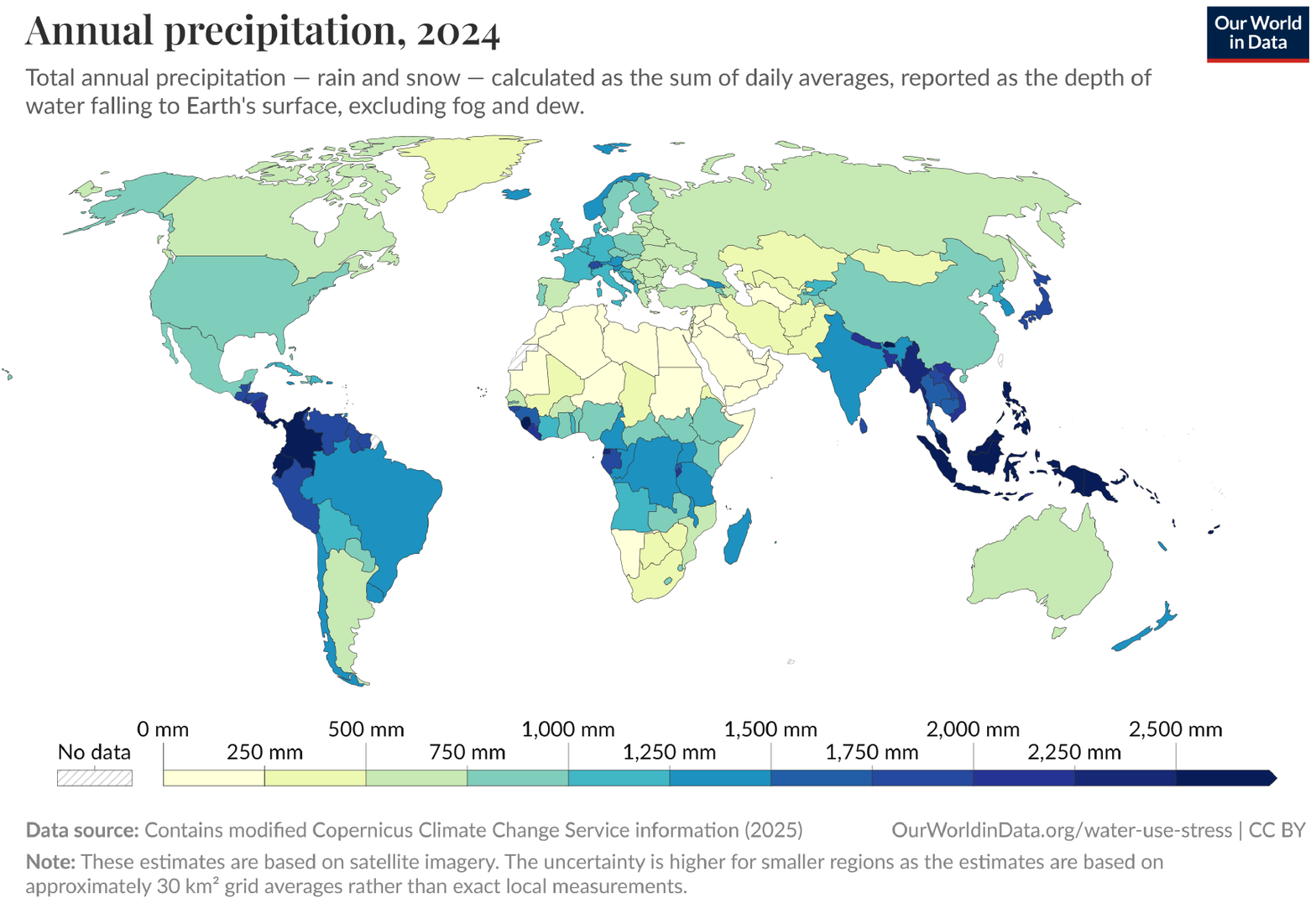}}
\caption{Total annual precipitation (rain and snow) calculated as the sum of daily averages in 2024. Image credit: [OurWorldinData].}
\label{fig:fig1}
\end{figure}
Table \ref{tab:TabI} lists the selected rain focused related works, publication year, and information about the experimental setup (terminal, link direction, protocols and latency) and the geographical region. Fig. \ref{fig:fig1} illustrates the total annual precipitation, including both rain and snow, which is calculated as the sum of daily averages. This could lead to substantial variations in Starlink throughput for the users across different regions. According to Starlink, it is designed to operate in rain, however,  heavy rain could affect the connection, resulting in lower throughput or a rare outage.

Starlink recommends using high performance and FHP terminals for users living in areas with extreme weather conditions. However, the existing studies \cite{Sami,Sami2,WetLinks,Careau} do not use these recommended terminals when examining the impact of rain.  FHP terminal measurements during both clear and rainy weather conditions are conducted in Germany \cite{Dominic_2} and the United States \cite{Bin}. However, these studies do not explicitly discuss and compare the performance during different weather conditions. Notably, the number of Starlink satellites passing over the Arctic region near Oulu, Finland, is significantly lower compared to visible satellites in Germany and the United States \cite{Mohan,Beckman}. Based on the literature review, there is a clear need to use the FHP terminal and conduct a more thorough bidirectional measurement campaign to study the impact of weather on Starlink performance.  To the best of the authors' knowledge, no previous studies have investigated the effects of weather on the performance of the Starlink FHP terminal near the Arctic region or Northern Europe.  Our paper addresses this shortcoming with FHP terminal measurements near the Arctic region to answer the following questions.

\begin{itemize}
    \item To what extent does Starlink's performance degrade during periods of rain and high humidity?
    \item Does rainfall lead to complete service outages with no connection, and if so, which link—uplink or downlink—is more susceptible to degradation due to precipitation?
    \item To what degree does the cloudy sky pose a significant challenge to the stability of Starlink throughput?
\end{itemize}

The rest of this paper is organized as follows. Section~\ref{sec:sec2} presents the Starlink technical specifications. In Section~\ref{sec:sec3} and Section~\ref{sec:sec4}, we discuss the experimental setup and collected dataset, respectively.  Section~\ref{sec:sec5} discusses the measurement results and the lessons learned. Finally, Section~\ref{sec:sec6} concludes this paper with final remarks.

\section{Technical Specifications}
\label{sec:sec2}
Starlink offers different terminals and service plans tailored to the needs of the users. The common use cases include stationary residential connection, mobile roaming for recreational vehicles and campers, and in-motion connectivity for vessels, vehicles and aircraft. Starlink recommends specific ground terminals matched with corresponding service plans for each use case to deliver optimal performance~\cite{Asad_Starlink}.
\subsubsection{Terminal}
At the time of writing this paper, Starlink offers five ground terminal options, including (i) Mini, (ii) standard, (iii) standard actuated, (iv) high performance, and (v) flat high performance.  All of these terminals use an electronic phased array (EPA) antenna and can operate in temperatures from -30$^\circ$C to  50$^\circ$. However, these terminals' dimensions, field-of-view, average power consumption, weight, and orientation capabilities vary as listed in Table \ref{tab:Terminals}. 

The standard and mini terminals feature software-assisted manual orienting (SAMO), which is manual alignment toward the sky using the Starlink smartphone application. The standard actuated and high performance terminals feature a motorized self-orienting (MSO) alignment, which automatically points the terminal toward the sky in an optimal direction. In this paper, we use the FHP terminal. Due to the fixed installation, the FHP terminal neither supports SAMO nor MSO alignment feature.

\begin{table*}[t!]
    \centering
    \caption{Starlink terminal and key specifications as of March 13, 2025.}
    \begin{tabularx}{\textwidth}{|l|>{\centering\arraybackslash}X|>{\centering\arraybackslash}X|>{\centering\arraybackslash}X|>{\centering\arraybackslash}X|>{\centering\arraybackslash}X|}
        \hline
        \textbf{Functions} & \textbf{Standard} & \textbf{Standard} & \textbf{High} & \textbf{Flat High} & \textbf{Mini} \\
        & \textbf{} & \textbf{Actuated} & \textbf{Performance} & \textbf{ Performance} & \textbf{} \\
        \hline
      \textbf{Antenna} & EPA & EPA & EPA & EPA & EPA \\
       \textbf{dimension (mm $\times$ mm)} & 594 $\times$ 383 & 513 $\times$ 303 & 575 $\times$ 511& 575 $\times$ 511 & 298.5 $\times$ 259 \\
       \textbf{Field of view} & 110$^\circ$ & 100$^\circ$ & 140$^\circ$ & 140$^\circ$ & 110$^\circ$ \\
        \textbf{Orientation} & SAMO & MSO & MSO & Fixed & SAMO\\
        \textbf{Average power usage} & 75-100 W& 50-75 W &  110-150 W& 110-150 W &  25-40 W\\
        \textbf{Operating temperature} & -30$^\circ$C to 50$^\circ$C & -30$^\circ$C to 50$^\circ$C & -30$^\circ$C to 50$^\circ$C & -30$^\circ$C to 50$^\circ$C& -30$^\circ$C to 50$^\circ$C\\
      \textbf{Snow melt capability} & \multirow{1}{*}{40 mm/h}&  \multirow{1}{*}{40 mm/h} & \multirow{1}{*}{75 mm/h}  & \multirow{1}{*}{75 mm/h}  &  \multirow{1}{*}{25 mm/h} \\

    \textbf{Wind rating} & 96+ km/h & 80+ km/h & 80+ km/h & 280+ km/h & 96+ km/h \\
\textbf{In-motion capability} &  Recommended &  &  & Recommended &  Recommended\\
\textbf{Extreme weather conditions} &  &   & Recommended &Recommended  & \\
        \hline
    \end{tabularx}
    \label{tab:Terminals}

\end{table*}

\subsubsection{Service Plan}
Starlink recommends the Standard service plan for households and the Priority plan for businesses and high-demand users. However, neither of these plans supports mobility. In contrast, the Mobile plan allows roaming, enabling users to move terminals between locations, making it suitable for RVs, nomads, and campers. The Mobile Priority (mobility) plan provides in-motion connectivity for both land vehicles and vessels at sea. This plan is particularly suited for maritime operations, emergency response, and mobile businesses. In this paper, the terminal is subscribed to a Mobile Priority (mobility) service plan.

\begin{figure}[t!]
\centerline{\includegraphics*[width=0.5\textwidth]{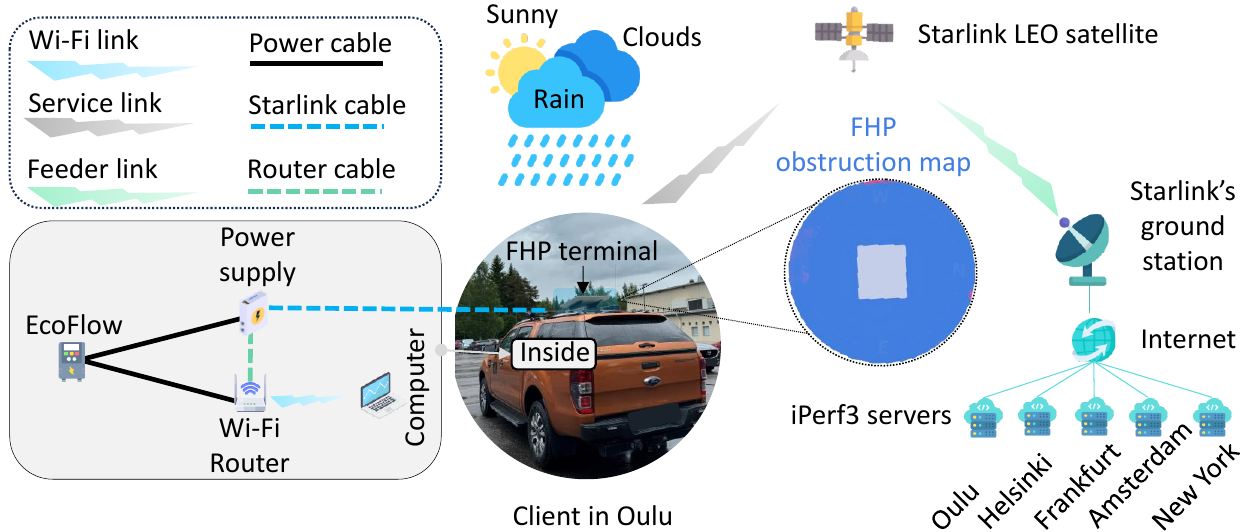}}
\caption{Experimental setup illustration, where flat high performance terminal is installed on the rooftop of a Ford Ranger vehicle.}
\label{fig:fig2}
\end{figure}
\section{Experimental Setup}
\label{sec:sec3}

In this section, we discuss the measurement testbed as illustrated in Fig. \ref{fig:fig2}\footnote{Icons are made by multiple authors from www.flaticon.com including freepik, Vectorslab, juicy\_fish and Design Circle.}. Table \ref{tab:TabII} provides the information about used terminal, service plan, measurement location, servers, and days.

\subsubsection{Hardware}
We installed the Starlink's FHP terminal on the rooftop of the Ford Ranger vehicle using Wedge Mount, which has 8$^\circ$ tilt to facilitate water runoff during the rain. For measurements, we used a laptop with specifications as Intel(R) Core(TM) i5-6200U CPU operating at 2.30 GHz, and it featured an Intel(R) Dual Band Wireless-AC 8260 Wi-Fi adapter. We connected the laptop with the Starlink router using Wi-Fi connection. Refer to Section III of \cite{Asad_Starlink} which discusses the detailed hardware setup.
\begin{table}[t]
    \centering
    \caption{Information of used terminal, service plan, measurement location, servers and days.}
    \resizebox{\columnwidth}{!}{\begin{tabular}{|l|c|}
        \hline
        \textbf{Feature} & \textbf{Description}\\
        \hline
        \textbf{Starlink terminal} & flat high performance\\
        \textbf{Terminal installation} & Vehicle rooftop\\
        \textbf{Service plan} & Mobile priority (mobility)\\
        \multirow{2}{*}{\textbf{Client location}} & Latitude: 65.05751426732301\\
        &  Longitude: 25.456422139689952\\
        \multirow{2}{*}{\textbf{Server location}} & Oulu, Helsinki, Frankfurt, \\
        &Amsterdam, New York\\
    \textbf{Measurement scenario} & Stationary\\
    \textbf{Measurement days} & June 12, 18, 20, 26, and 27, 2024\\
     \textbf{Measurement tools} & Ping, iPerf3, and Traceroute\\
    \textbf{Sampling rate} & 1 second\\
        \hline 
    \end{tabular}}
    \label{tab:TabII}
\end{table}
\subsubsection{Software}
To be consistent with the earlier works~\cite{Sami,Dominic_2024,Bin,WetLinks,Kassem,Asad_Starlink}, we used the following \verb|iperf3|, \verb|ping| and \verb|tracert| tools for throughput, RTT, and hops count measurements, respectively.

\begin{itemize}
\item \textbf{iPerf3}   supports multiple parallel traffic streams simultaneously, to measures the maximum achievable throughput. In our measurement, we used TCP for uplink and downlink with ten parallel streams. We used the \verb|iperf3| version 3.17 and a default configuration, including TCP congestion control.
  \item \textbf{Ping}  We used the Windows default \verb|ping| setting to measure RTT.  A single \verb|ping| test sends a 32 bytes packet to a server and measures the time a packet takes to go from the client to the server and back to the client.
  \item \textbf{Traceroute} We used \verb|tracert| to understand how the uplink data travels from the client to the server.
\end{itemize}

\subsection{Client and Server Locations}
\subsubsection{Client}
We parked the vehicle carrying the Starlink FHP terminal in the parking area of the VTT Technical Research Center Ltd, Oulu, approximately 9 km away from Oulu city. There were no blockages in the surrounding area, which was also confirmed by the Obstruction Map generated by the Starlink application, as shown in Fig. \ref{fig:fig2} and Fig. \ref{fig:locations}.
\subsubsection{Server}
We used the Hostkey \verb|iperf3| servers located in Helsinki, Amsterdam, Frankfurt and New York\footnote{Hostkey B.V. Speedtest (https://speedtest.hostkey.com/)}.  Additionally, we set up a fifth \verb|iperf3| server in  Oulu, Finland. \verb|iperf3| server was running on a computer with the \verb|Ubuntu 22.02| operating system inside VTT’s test network\footnote{5G and 6G test network environment by VTT (\url{https://www.vttresearch.com/en/5g-and-6g-test-network-environment})}. Due to security reasons, we are not providing the \verb|hostname| for our \verb|iperf3| server in Oulu. Notably, Starlink offers 100 \% coverage to these locations\footnote{Starlink Availability Map (\url{https://www.starlink.com/map})}. Table \ref{tab:tabIII} lists the servers' \verb|hostname|.

\begin{table}[!t]
\caption{The hostname of examined iPerf3 servers.\label{tab:tabIII}}
\centering
  \resizebox{\columnwidth}{!}{\begin{tabular}{|l|c|c|c|}
\hline
No. & Server location & Hostname &Owner\\
\hline
1& Helsinki, Finland & spd-fisrv.hostkey.com & Hostkey\\
2.& Amsterdam, Netherlands& spd-nlsrv.hostkey.com & Hostkey\\
3.& Frankfurt, Germany& spd-desrv.hostkey.com & Hostkey\\
4.& New York, United States& spd-uswb.hostkey.com & Hostkey\\
5.& Oulu, Finland & Anonymous & VTT\\
\hline
\end{tabular}}
\end{table}

\section{Measurement Dataset}
\label{sec:sec4}
\begin{figure}[t!]
\centerline{\includegraphics*[width=0.5\textwidth]{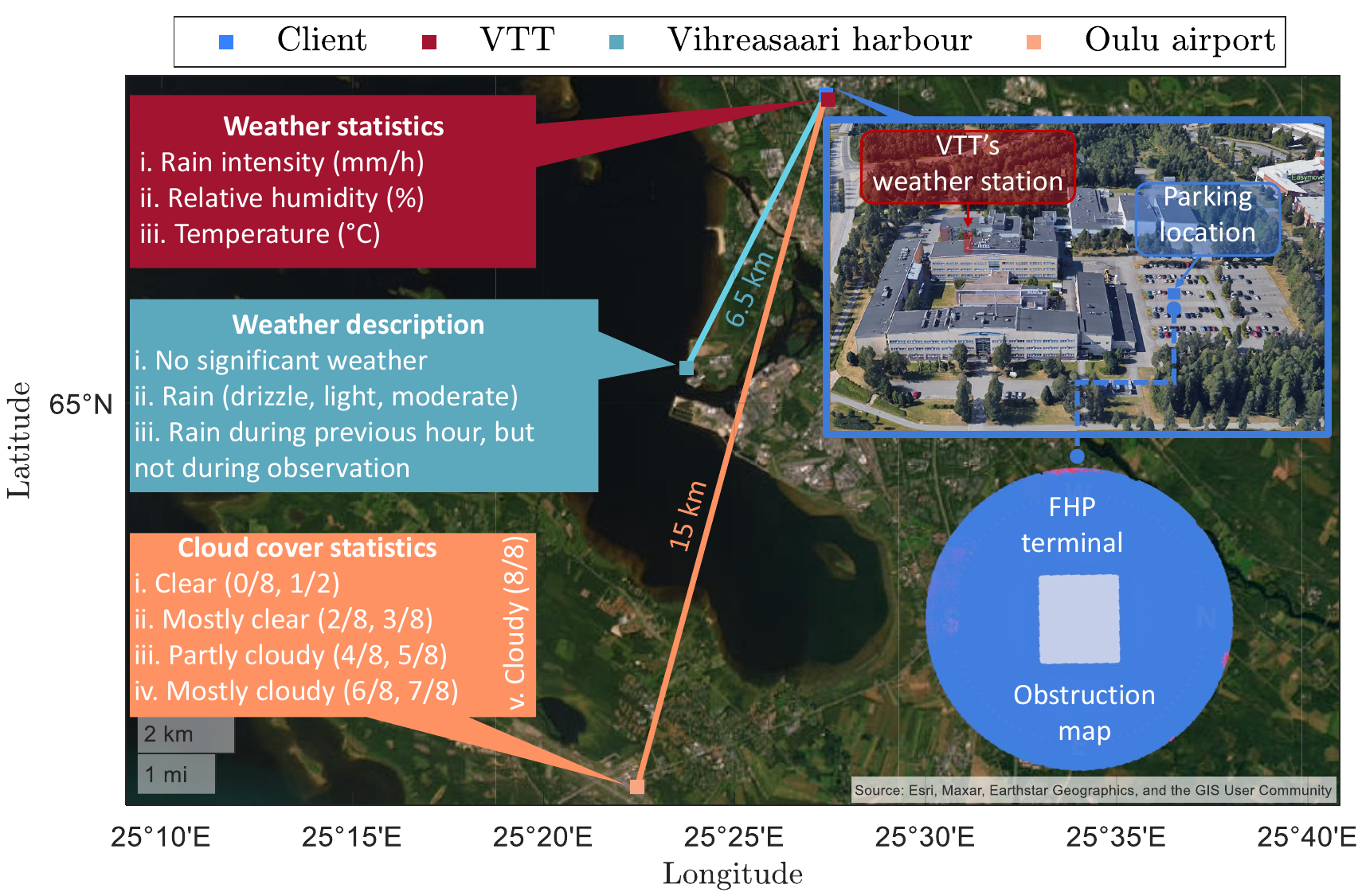}}
\caption{Location of the measurement (client) and the weather stations.}
\label{fig:locations}
\end{figure}

\subsection{Weather Statistics}
To complement our analysis, we used three different weather stations within Oulu, which provide meteorological statistics at the sampling rate of 1 minute. Fig. \ref{fig:locations} shows the location of the measurement and the weather stations. First, we used VTT Willab's weather station in Linnanmaa, Oulu\footnote{Weather in Oulu, Linnanmaa (https://weather.willab.fi/weather.html)}, to access the meteorological statistics of relative humidity, temperature, and rain intensity. Notably, the client measurement location was only a few meters apart from this weather station as illustrated in Fig. \ref{fig:locations}. Second, we used the Finnish Meteorological Institute~(FMI)\footnote{Finnish Meteorological Institute (https://www.ilmatieteenlaitos.fi/saa/oulu)}  weather station in Vihreäsaari harbour  to determine the weather description (e.g., i. No significant weather, ii. Rain during the previous hour but not during observation, and iii. Rain). The shortest distance (displacement) between the Vihreäsaari harbour weather station and the measurement location is around 6.5~km. It is worth emphasizing that neither VTT's nor Vihreäsaari harbour weather station provides information about the clouds cover. Therefore, we leveraged the FMI's Oulu airport as a third weather station source to obtain cloud cover information. The displacement between FMI's Oulu airport weather station and the VTT, Oulu, is around 15~km. By considering the Starlink's spot beam's diameter of 25 km and unknown elevation angle, we assume that these three weather stations and the client's location experienced similar weather conditions during the measurements. Additionally, we cross-checked these three different datasets, and there were similarities during the measurement period. 
\subsection{Starlink Measurements}
\begin{table}[!t]
\caption{Measurement date, weather description and number of collected samples.}
\label{tab:tabIV}
\centering
  \resizebox{\columnwidth}{!}{\begin{tabular}{|l|c|c|c|c|}
\hline
No. & Weather description& Date &UL & DL\\
\hline
\multirow{2}{*}{1.} &No significant weather & \multirow{2}{*}{27.06.24} & \multirow{2}{*}{3,119}  & \multirow{2}{*}{3,161}\\
 &Cloud cover (Clear) &  & &  \\
 \hline
\multirow{2}{*}{2.} &No significant weather & \multirow{2}{*}{26.06.24} & \multirow{2}{*}{3,690}  & \multirow{2}{*}{3,690} \\
 &Cloud cover (Mostly Cloudy) &  & & \\
 \hline
\multirow{2}{*}{3.} &Rain during previous hour & \multirow{2}{*}{12.06.24} &  \multirow{2}{*}{730}& \multirow{2}{*}{600}  \\
& but not during observation &  &  &   \\
\hline
\multirow{2}{*}{4.}& \multirow{2}{*}{Rain (drizzle, light, moderate)}  & 12.06.24 & \multirow{2}{*}{2,277}  & \multirow{2}{*}{2,761}  \\
& & 18.06.24 &   &   \\
\hline
\end{tabular}}
\end{table}
We conducted measurements with a sampling rate of 1 second. During our measurements, we collected 9,816 and 10,212 measurement samples for uplink and downlink, respectively.  Table \ref{tab:tabIV} shows the weather description,  measurement dates, and number of uplink (UL) and downlink (DL) samples. In \cite{Asad_Starlink}, the \verb|iperf3| measurement results confirmed that the server location has no significant impact on throughput. Therefore, in this paper, we do not distinguish the \verb|iperf3| measurements by server location and instead analyze the performance of all servers collectively. Additionally, we collected 2,031 RTT samples for the server in Oulu and Amsterdam.  Unlike \verb|iperf3| throughput, we separately examine the RTT for these two servers,  We conducted the measurements on June 12, June 18, June 26, and June 27, 2024. On June 12 and June 18, the measurements were conducted during the rain (drizzing, light to moderate) and also within one hour after the rain (which experienced high humidity). On June 26, the percentage of cloud cover was up to 87.5\%. On June 27, there were no clouds, and the sky was clear.
\section{Measurement results}
\label{sec:sec5}
\subsection{Impact of Rain}
\begin{figure}[t!]
\centerline{\includegraphics*[width=0.5\textwidth]{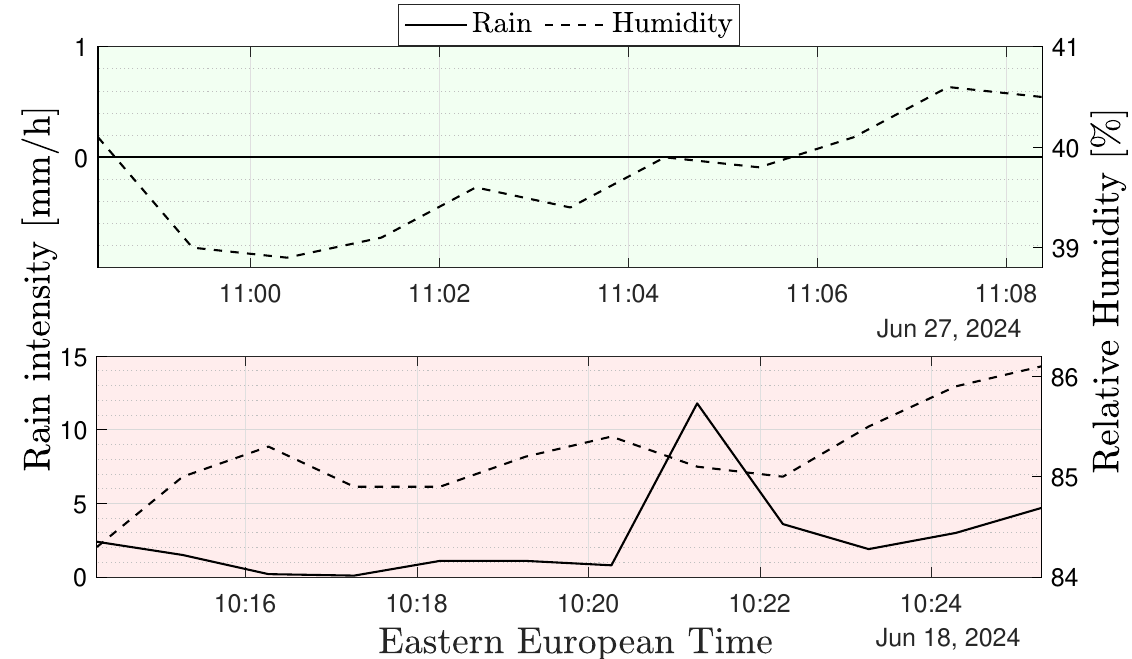}}
\caption{Comparison of rain intensity and relative humidity for June 18 and June 27.  The upper plot (green) shows that on June 27, there was no rain, the sky was clear without cloud cover, and relative humidity was almost half. The lower plot (red) of the figure shows statistics for June 18; there was rain, causing high relative humidity. The sampling rate of weather statistics is 60 seconds.}
\label{fig:fig3}
\end{figure}

\begin{figure}[t!]
\centerline{\includegraphics*[width=0.5\textwidth]{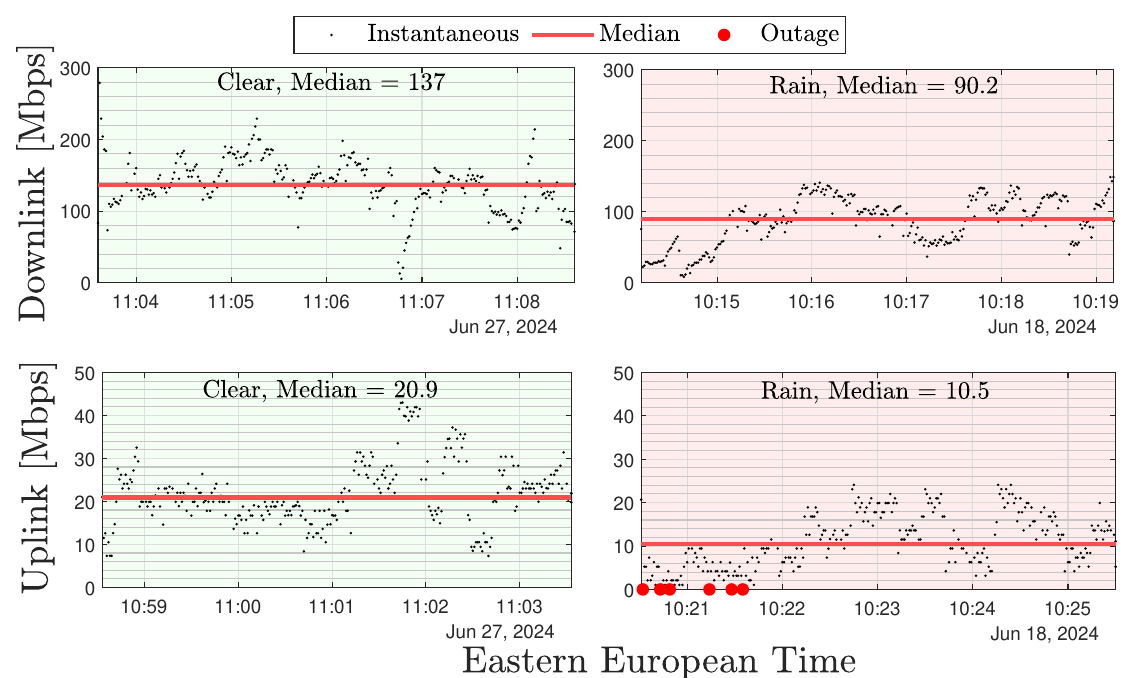}}
\caption{Throughput comparison of a clear and rainy day from an iPerf3 client in Oulu to an iPerf3 server in Helsinki. Each sub-figure contains 300 samples corresponding to a 5-minute measurement period. In total, this figure comprises 1,200 samples.}
\label{fig:fig4}
\end{figure}

Fig. \ref{fig:fig3} shows the relative humidity and rain intensity for ten minutes internals on two different days, June 18, 2024, and June 27, 2024. On June 18, it was raining with intensity up to 12.5 mm/h, where the relative humidity was over 85\%.  On the contrary, there was no rain or clouds on June 27, and the relative humidity was almost half compared to June 18. Fig.~\ref{fig:fig3} lays the foundation to understand the impact of weather conditions on Starlink's uplink and downlink performance in Fig.~\ref{fig:fig4}. Additionally, the temperature was around 17$^\circ$C and 20$^\circ$C on June 18 and June 27, respectively.

Fig. \ref{fig:fig4} reveals the impact of rain on downlink and uplink throughput. One can observe the median downlink throughput drops from 137 Mbps to 90.2 Mbps during the rain. Similarly, the median uplink declined from 20.9 Mbps to 10.5 Mbps. Notably, there were six outages, each lasting one second, during uplink measurements in the rain between 10:20 and 10:22 on June 18, 2024. In Fig. \ref{fig:fig3}, one can see that the rain intensity was up to 12.5 mm/h during the same time, and Vihreäsaari harbour weather stations characterized it as moderate rain. This explains the reason behind these outages and the drop in throughput.

\begin{figure}[t!]
\centerline{\includegraphics*[width=0.5\textwidth]{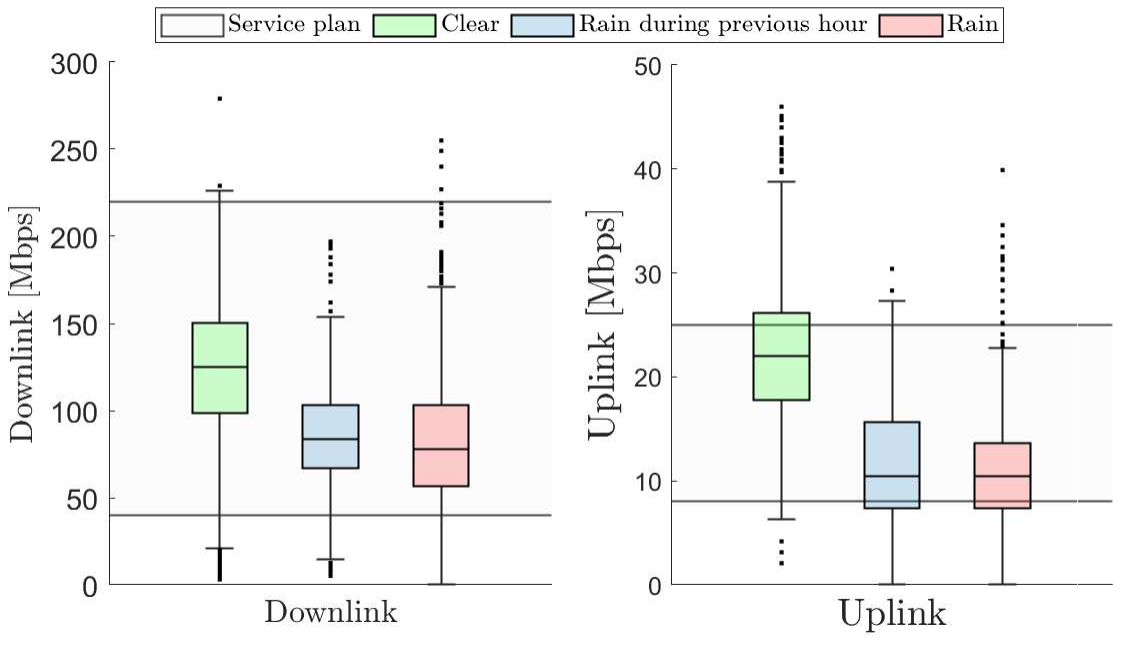}}
\caption{Throughput observation for different weather conditions including clear day, after the rain, and during rain. In this figure, the downlink dataset has 3,161, 600 and 2,761 samples for clear day, after the rain, and during rain, respectively.  Uplink dataset comprises 3,119, 730 and 2,277 samples for clear day, after the rain, and during rain, respectively. These results include all five servers mentioned in Table~\ref{tab:tabIII}.}
\label{fig:fig5}
\end{figure}

Fig. \ref{fig:fig5} uses boxchart\footnote{Box chart (https://se.mathworks.com/help/matlab/ref/boxchart.html)} and examines a large dataset given in Table \ref{tab:tabIV} and compares throughput of (i) service plan advertised by Starlink; (ii) no significant weather which means no rain; (iii) right after the rain within one hour which means high relative humidity; and (iv) rain including drizzle, light, moderate. 
Similar to the observations in Fig. \ref{fig:fig4}, one can see that median uplink and downlink decrease by 52.2\% and 35\%, respectively. Despite these significant performance degradations, the median throughput remains within the limits specified by Starlink.

\begin{figure}[t!]
\centerline{\includegraphics*[width=0.5\textwidth]{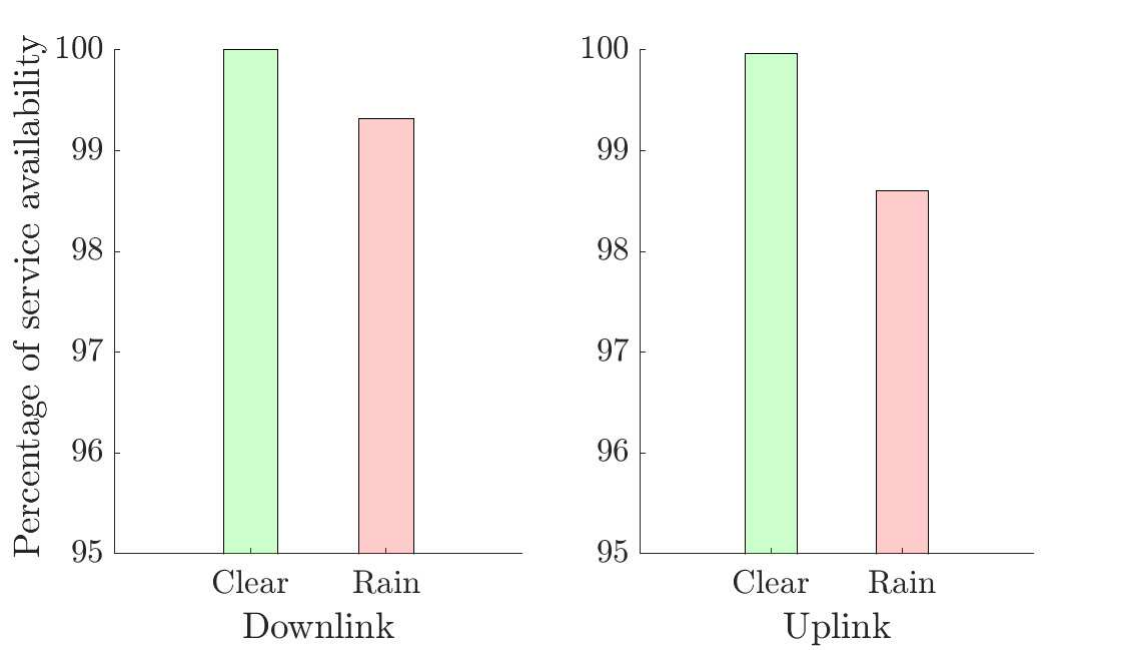}}
\caption{Comparison of service availability for downlink and uplink during clear and rainy weather. These results include all data samples from Fig. \ref{fig:fig5}. }
\label{fig:fig6}
\end{figure}

Fig. \ref{fig:fig6} shows the service availability calculated as a percentage of the time when \verb|iperf3| throughput was non-zero. Given the sampling rate of 1 second, each outage is at least one second long. Despite the rain and high relative humidity, the overall service availability remains above 98.5 for uplink and downlink. Notably, uplink has 1\% lower service availability than downlink.  This confirms that Starlink can maintain the connection during rain. However, the throughput decreases significantly, as illustrated in Fig. \ref{fig:fig4} and Fig. \ref{fig:fig5}. Notably, most of our measurements were conducted during light rain, we expect a lower service availability for moderate and heavy rain scenarios.

\begin{figure}[t!]
\centerline{\includegraphics*[width=0.5\textwidth]{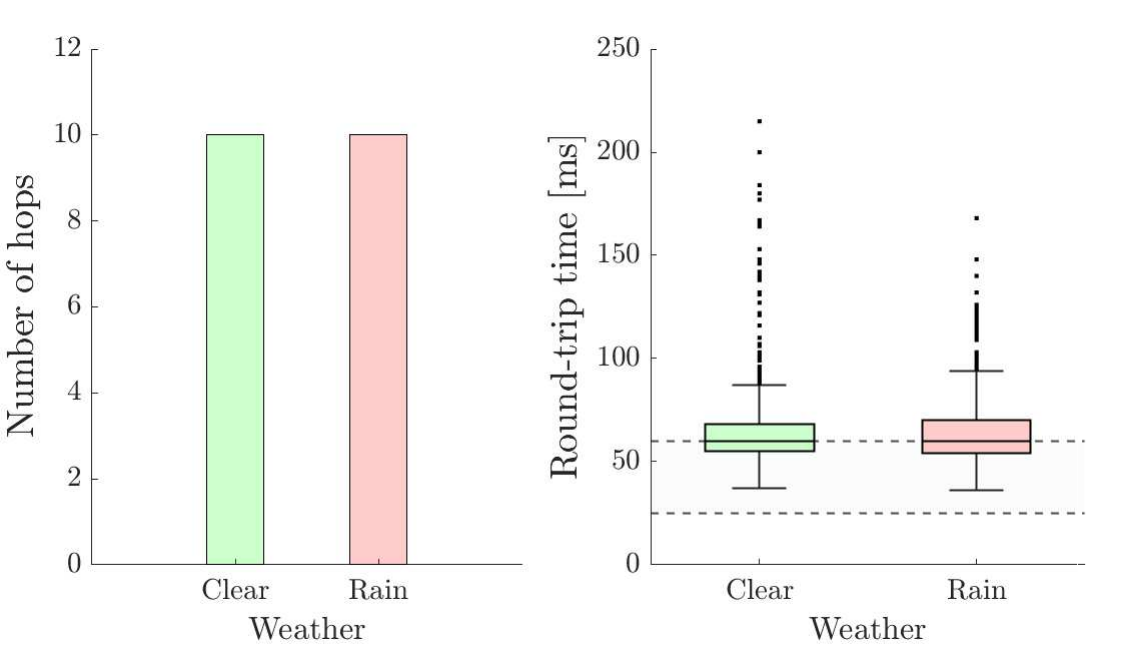}}
\caption{Comparison of the number of hops and RTT from a client in Oulu to a server in Amsterdam. RTT has a total of 1,354 samples, where each weather condition contributes 677 samples.}
\label{fig:fig7}
\end{figure}

Fig. \ref{fig:fig7} presents the number of hops and RTT between a client in Oulu to a server in Amsterdam. The number of hops and median RTT remains the same for the measurements with and without the rain. However, RTT measurements during rain have a larger maximum RTT value. 

\subsection{Impact of Clouds}
\begin{figure}[t!]
\centerline{\includegraphics*[width=0.5\textwidth]{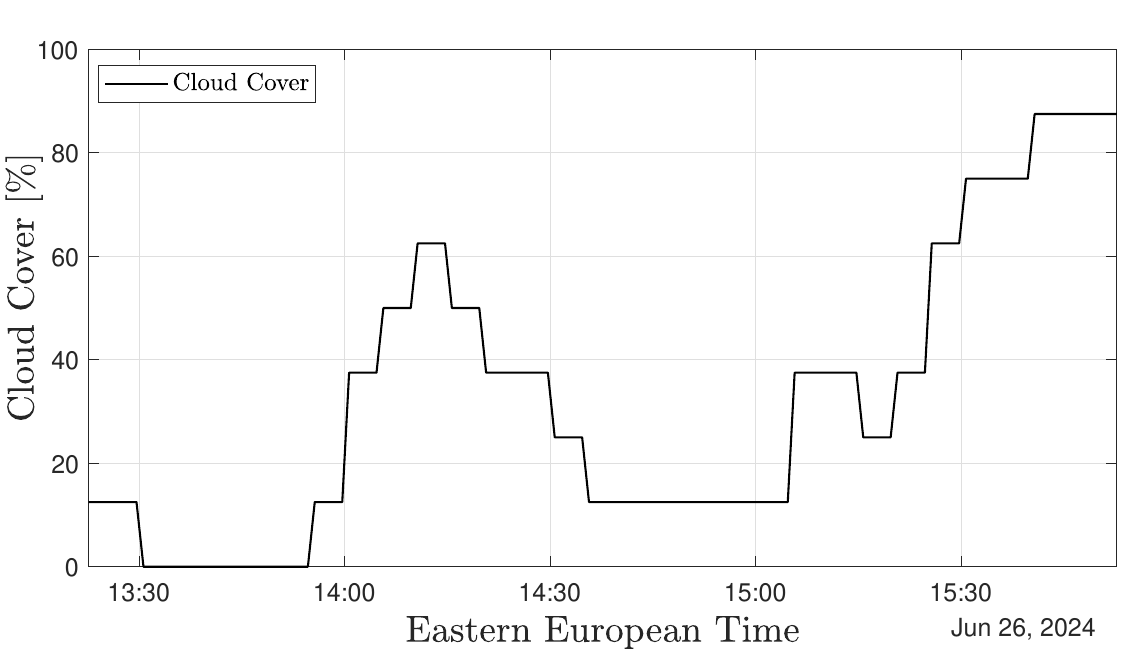}}
\caption{Percentage of the cloud cover during the June 26, 2024, measurements.}
\label{fig:fig8}
\end{figure}

It is worth to remind that Starlink uses Ku-band, which makes its signals vulnerable to cloud attenuation. This paper also study the impact of weather on the Starlink measurements. It is worth reminding that we collected the clouds cover information from the FMI's Oulu Airport weather station. Fig.~\ref{fig:fig8} shows the percentage of the cloud cover as a function of time from 13:22:48 to 15:51:05 on June 26, 2024. One can see that cloud cover reached 87.5\%, occupying most of the sky at the end of the measurements.

\begin{figure}[t!]
\centerline{\includegraphics*[width=0.5\textwidth]{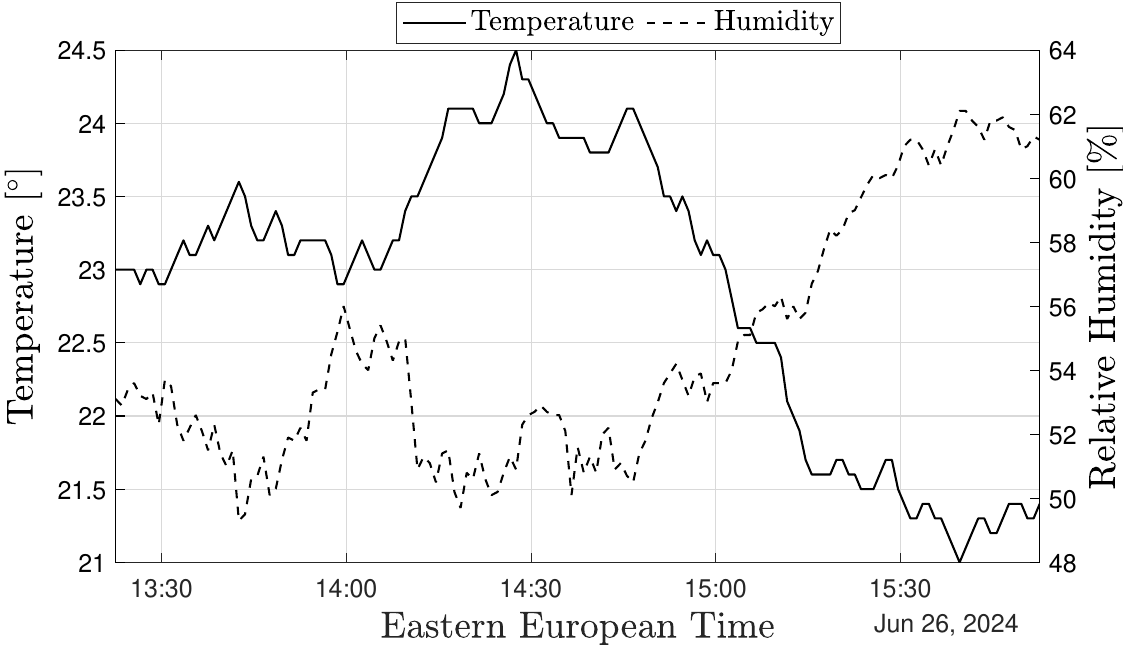}}
\caption{Impact of cloud cover on temperature and relative humidity during the June 26, 2024, measurements.}
\label{fig:fig9}
\end{figure}

To complement this cloud cover data, we used the relative humidity and temperature information from the VTT Willab's weather station. Fig. \ref{fig:fig9} shows relative humidity and temperature as a function of time between 13:22:48 and 15:51:05 on June 26, 2024.  These results align with Fig. \ref{fig:fig8}; one can observe that the humidity increases with cloud cover while temperature slightly drops by two degrees from 23$^\circ$C to 21$^\circ$C.

\begin{figure}[t!]
\centerline{\includegraphics*[width=1\columnwidth]{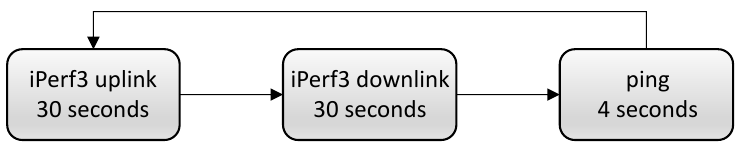}}
\caption{The process of evaluating Starlink's throughput and RTT from an iPerf3 client in Oulu to an iPerf3 server in Oulu during a cloudy day (June 26, 2024).}
\label{fig:fig10}
\end{figure}

To have more control on the server and conduct longer measurements, we used the server in Oulu for \verb|iperf3| and \verb|ping| tests. Fig. \ref{fig:fig10} shows the measurement process used to collect the uplink, downlink, and RTT samples on June 26, 2024. From 13:22:48 to 15:51:05, this process was 148 minutes long and was repeated 123 times for each metric. We collected 3,690, 3,690, and 492 samples for uplink, downlink, and RTT, respectively.

\begin{figure}[t!]
\centerline{\includegraphics*[width=1\columnwidth]{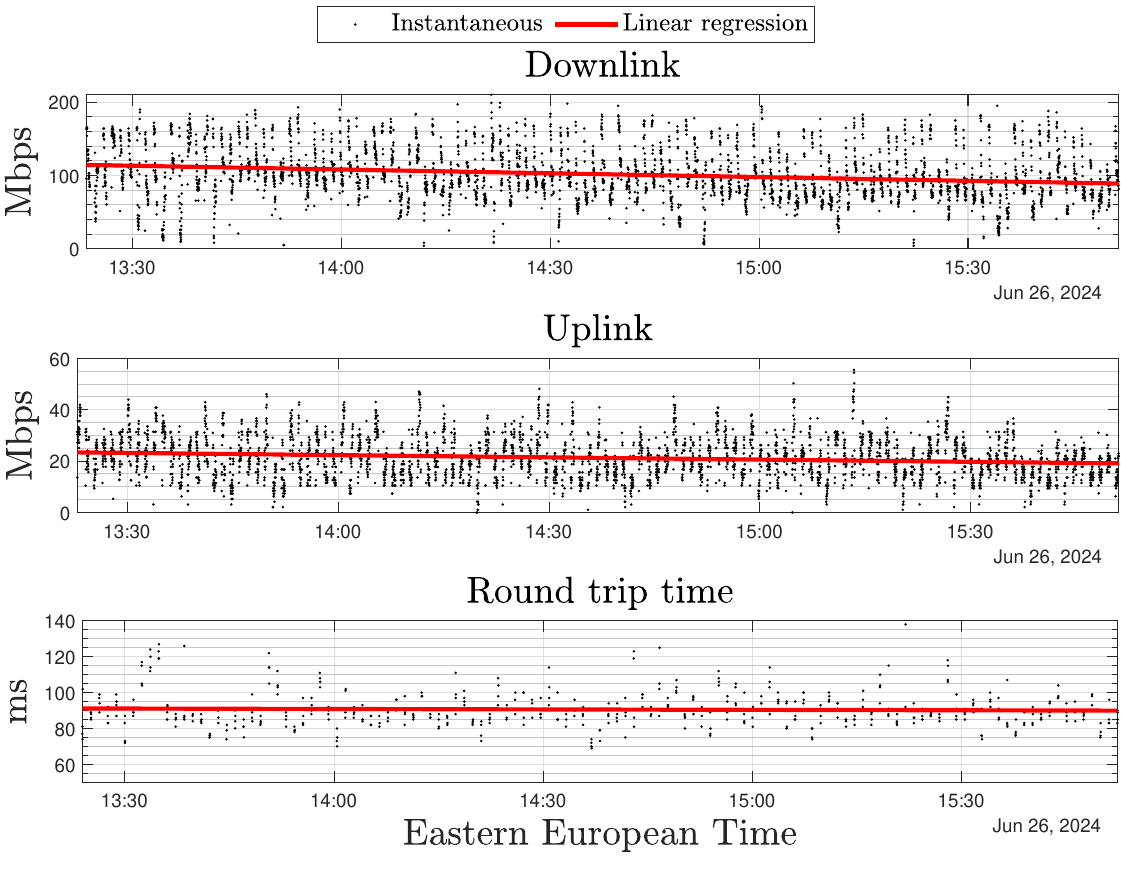}}
\caption{Throughput and RTT from an iPerf3 client in Oulu to an iPerf3 server in Oulu during a cloudy day. This figure includes a total of 7,380 samples, equally distributed between the downlink and uplink, with each consisting of 3,690 samples.  RTT contains 492 samples.}

\label{fig:fig11}
\end{figure}

Fig. \ref{fig:fig11} shows the measurements for the interval 13:22:48 to 15:51:05 on June 26, 2024. During this period, the cloud cover was up to 87.5\%, and relative humidity exceeded 60\%. With an increase in cloud cover and relative humidity, the linear regression of the collected throughput data samples allowed us to observe a decreasing trend for both uplink and downlink. Particularly, a cloud cover of up to 12.5\%  demonstrates approximately 20\%  higher throughput compared to the weather conditions with 87.5\% cloud cover. However, the linear regression of RTT remains stable throughout the measurement period.  This RTT observation aligns with the findings of \cite{Dominic_2}. 

\newpage
\section{Conclusion}
\label{sec:sec6}
Starlink uses Ku-band for uplink and downlink communication between the user and the satellite. This paper examines the impact of rain and cloud cover on the Starlink bidirectional throughput and RTT. Our results reveal that even a light rain significantly reduces the throughput both for uplink and downlink. The moderate rain caused connection losses, leading to several outages, each lasting one second. The
linear regression analysis reveals a negative relationship between throughput and cloud cover. Notably, we observed that the light rain and clouds do not impact the RTT. Moderate to heavy rainfall leads to connection loss, we expect such extreme weather conditions may contribute to higher RTT. However, further measurements and analysis are needed to investigate and understand how the rainfall intensity affects RTT. In future work, we plan to expand our measurement campaign for a longer duration to gather additional data, enabling a more comprehensive analysis of Starlink performance under extreme weather conditions, including storms, heavy rainfall, and snowfall.

\vspace{-5pt}
\section*{Acknowledgment}
This research was supported by Business Finland through the Drolo II and 6G-SatMTC projects; in part by the European Union’s Horizon Europe Smart Networks and Services Joint Undertaking (SNS JU) under Amazing 6G project (grant agreement No 101192035).
\vspace{12pt}
\bibliographystyle{IEEEtran}
\bibliography{IEEEabrv,ref}

\end{document}